\pdfoutput=1
\documentclass[pdftex]{pasj00}

\newcommand{\kt}{\ensuremath{k_{\rm{B}}T}}
\newcommand{\lx}{\ensuremath{L_{\rm{X}}}}

\newcommand{\fx}{\ensuremath{F_{\rm{X}}}}
\newcommand{\nh}{\ensuremath{N_{\rm H}}}

\SetRunningHead{D.~Takei et al.}{Detection of an X-Ray Transient Source}
\Received{2007/04/04}
\Accepted{2007/05/16}
\Published{}

\begin{document}

\title{Detection of a Rare Supersoft Outburst Event \\ during a Suzaku Observation of 1E\,0102.2--7219}

\author{
Dai \textsc{Takei}\altaffilmark{1},
Masahiro \textsc{Tsujimoto}\altaffilmark{1,2,3},
Shunji \textsc{Kitamoto}\altaffilmark{1},
Mikio \textsc{Morii}\altaffilmark{1},\\
Ken \textsc{Ebisawa}\altaffilmark{4},
Yoshitomo \textsc{Maeda}\altaffilmark{4},
and Eric D. \textsc{Miller}\altaffilmark{5}
}

\altaffiltext{1}{Department of Physics, Rikkyo University\\
3-34-1 Nishi-Ikebukuro, Toshima, Tokyo 171-8501}
\email{takei@stu.rikkyo.ne.jp}
\altaffiltext{2}{Department of Astronomy and Astrophysics, Pennsylvania State University\\
525 Davey Laboratory, University Park, PA 16802, USA}
\altaffiltext{3}{Chandra Fellow}
\altaffiltext{4}{Institute of Space and Astronautical Science, Japan Aerospace Exploration Agency\\
3-1-1 Yoshinodai, Sagamihara, Kanagawa 229-8510}
\altaffiltext{5}{Massachusetts Institute of Technology, Kavli Institute for Astrophysics and Space Research\\
77 Massachusetts Avenue 37-551, Cambridge, MA 02139, USA}


\maketitle

\begin{abstract}
 We report the detection of a transient X-ray source toward the Small Magellanic Cloud
 (SMC) using the X-ray Imaging Spectrometer (XIS) onboard the Suzaku telescope. The
 source was detected at the edge of the XIS image during a routine observation of the
 calibration source 1E\,0102.2--7219, a supernova remnant in the SMC. We constrained the
 source position using ray-tracing simulations. No such transient source was found at
 the position in the other Suzaku observations nor in all the available archived images
 of other X-ray missions for the last $\sim$28 years. The XIS spectrum can be explained
 by a single blackbody with a temperature of $\sim$72~eV, and an interstellar extinction
 of $\sim$4.9$\times$10$^{20}$~H~atoms~cm$^{-2}$ consistent with the value to the
 SMC. An additional absorption edge at $\sim$0.74~keV was also confirmed, which is
 presumably due to the absorption by helium-like oxygen ions. Assuming that the source
 is at the distance of the SMC, the X-ray luminosity in the 0.2--2.0~keV band is
 $\sim$10$^{37}$~erg~s$^{-1}$ and the radius of the source is $\sim$10$^{8}$~cm. The XIS
 light curve shows about a two-fold decline in X-ray flux during the 24~ks
 observation. Together with the archived data, the X-ray flux in the burst is at least
 three orders of magnitude brighter than the undetected quiescent level. All these
 properties are often seen among supersoft sources (SSSs). We conclude that the
 transient source is another example of SSS in the SMC.
\end{abstract}

\section{Introduction}
1E\,0102.2--7219 is the second brightest X-ray source in the Small Magellanic Cloud
(SMC) at a distance of $\sim$60~kpc. It was discovered by the Einstein satellite
\citep{seward81} and was found to be a shell-type supernova remnant with a radius of
$\sim$14\arcsec~\citep{gaetz00}. The source is particularly suited for the routine
calibration of the quantum efficiency, the energy gain and resolution of X-ray detectors
because of its soft and line-dominated spectrum, non-variable flux, and good visibility
from satellites throughout the year. Therefore, a large number of pointed observations
of this object was conducted by X-ray satellites including Einstein, ROSAT, ASCA,
Beppo-SAX, Chandra, XMM-Newton, and Suzaku (\cite{hayashi94}; \cite{gaetz00};
\cite{hughes00}; \cite{sasaki01}; \cite{rasmussen01}; \cite{flanagan04};
\cite{sasaki06}). This offers a unique opportunity to search for transient sources in
the surrounding area and monitor their long-term behavior without additional investments
of telescope times.

Among sixteen Suzaku observations conducted by 2007 March, we detected a transient
source at the edge of the image obtained on 2005 August 31. As we discuss in this paper,
no such transient source was detected in any of the remaining fifteen Suzaku and
archived Einstein, ROSAT, ASCA, Beppo-SAX, Chandra, and XMM-Newton observations despite
their frequent visits, which indicates that the burst event of this source is quite
rare. Here, we report the result of the Suzaku observation of the transient source and
discuss its nature.

\section{Observations}
Sixteen observations of 1E\,0102.2--7219 were performed using the Suzaku satellite by
2007 March (table~\ref{tb:t1}). Suzaku \citep{mitsuda07} has two instruments in
operation; the X-ray Imaging Spectrometer (XIS; \cite{koyama07}) and the Hard X-ray
Detector (HXD; \cite{takahashi07}; \cite{kokubun07}). We concentrate on the XIS data in
this paper.

The XIS is equipped with four X-ray charge coupled devices (CCDs) at the foci of four
X-Ray Telescopes (XRT; \cite{serlemitsos07}). Three of them (XIS0, 2, and 3) are
front-illuminated (FI) CCDs sensitive in the 0.4--12~keV energy range and the remaining
one (XIS1) is a back-illuminated (BI) CCD sensitive in 0.2--12~keV. The total effective
area is $\sim$1360~cm$^{2}$ at 1.5~keV. XIS covers a $\sim$18\arcmin$\times$18\arcmin\
field of view (FoV), with an energy-independent half power diameter of
1\farcm8--2\farcm3. Each CCD has a format of 1024$\times$1024 pixels with a pixel scale
of $\sim$1\arcsec~pixel$^{-1}$. Two radioactive \atom{Fe}{}{55} sources illuminate two
corners of each CCD.

The energy resolution and quantum efficiency of the XIS are gradually degrading in
orbit. Their trend in the soft band is monitored using 1E\,0102.2--7219 in the same data
sets presented here. As of 2005 August 31, when the transient source was detected, the
absolute energy scale is accurate to $\lesssim$5~eV, and the energy resolution is
$\sim$63~eV (FI) and $\sim$71~eV (BI) at 1.0~keV. Due to an unknown contaminant
accumulating on the optical blocking filters of the XIS, the effective area in the soft
band has been diminishing since shortly after launch \citep{koyama07}. This effect is
accounted for in the ancillary response files and ray-tracing simulators of the
telescopes \citep{ishisaki07}.

All the XIS data were taken with the normal clocking mode with a frame time of 8~s. The
first two observations were aimed at (R.\,A., Dec.)\,$=$\,(\timeform{01h04m02s},
\timeform{-72D02'00''}), while the others were at (R.\,A.,
Dec.)\,$=$\,(\timeform{01h04m02s}, \timeform{-72D01'53''}) in the equinox J2000.0. We
used the data products of the processing version 1.2, in which events were removed
during the South Atlantic anomaly passages, the night earth elevation angle below
5~degrees, or the day earth elevation angle below 20~degrees \citep{fujimoto07}.

\begin{table}
 \begin{center}
 \caption{Suzaku observation log}\label{tb:t1}
  \begin{tabular}{cllcc}
   \hline\hline
   \phantom{000} & \multicolumn{1}{c}{Sequence} & \multicolumn{1}{c}{Start} & $\textit{t}\rm{_{exp}}$\footnotemark[$*$] & \phantom{000} \\
   & \multicolumn{1}{c}{number}   & \multicolumn{1}{c}{date}  & (ks) & \\
   \hline
   & 100001020                         & 2005-08-13 & \phantom{0}4 & \\
   & 100014010\footnotemark[$\dagger$] & 2005-08-31 & 24           & \\
   & 100044010                         & 2005-12-16 & 71           & \\
   & 100044020                         & 2006-01-17 & 42           & \\
   & 100044030                         & 2006-02-02 & 21           & \\
   & 101005010                         & 2006-04-16 & 22           & \\
   & 101005020                         & 2006-05-21 & 19           & \\
   & 101005030                         & 2006-06-26 & 22           & \\
   & 101005040                         & 2006-07-17 & 22           & \\
   & 101005050                         & 2006-08-25 & 49           & \\
   & 101005060                         & 2006-09-19 & 11           & \\
   & 101005070                         & 2006-10-21 & 37           & \\
   & 101005090                         & 2006-12-13 & 28           & \\
   & 101005100                         & 2007-01-15 & 24           & \\
   & 101005110                         & 2007-02-10 & 36           & \\
   & 101005120                         & 2007-03-18 & 18           & \\
   \hline
   \multicolumn{5}{@{}l@{}}{\hbox to 0pt{\parbox{80mm}{\footnotesize
	\footnotemark[$*$] Averaged exposure time of the operating CCDs.
        \par\noindent
       	\footnotemark[$\dagger$] The transient source was detected in this observation.
   }\hss}}
  \end{tabular}
 \end{center}
\end{table}

\section{Analysis}
\subsection{Image Analysis}
\begin{figure}
 \begin{center}
  \FigureFile(80mm,80mm){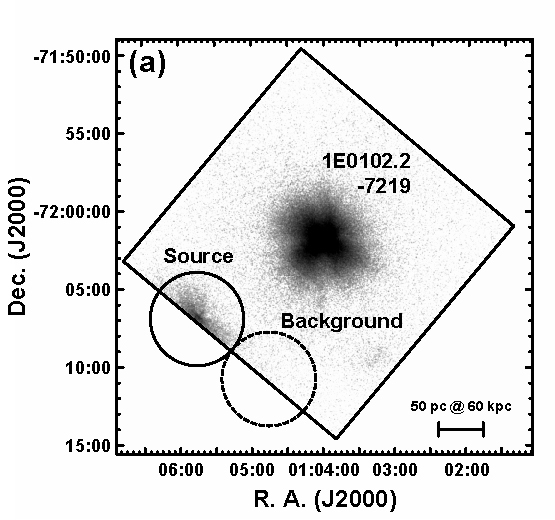}
  \FigureFile(80mm,80mm){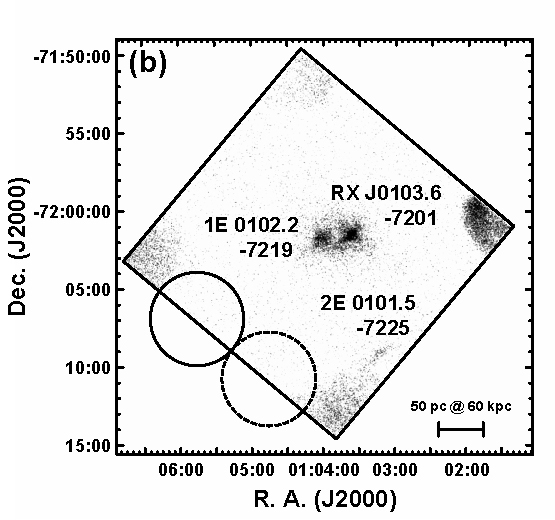}
 \end{center}
 \caption{
 XIS images in the (a) 0.2--2.0~keV and (b) 2.0--12~keV bands. Events taken with the
 four XIS were merged in the data on 2005 August 31. The bright spots in the four
 corners in (b) are the calibration sources. The solid and the dashed circles indicate
 the source and background accumulation regions, respectively. The names of the three
 persistent sources are given. The XIS frame was registered to match the observed
 position of 1E\,0102.2--7219 with the Chandra position.}\label{fg:f1}
\end{figure}

Figure~\ref{fg:f1} shows the XIS image obtained on 2005 August 31. Events taken with the
four detectors were merged. XIS detected four sources. Three of them were identified as
1E\,0102.2--7219, RX\,J0103.6--7201, and 2E\,0101.5--7225 using the SIMBAD
database. These sources were detected in all images taken at the other epochs. The
remaining one is a transient source detected only in the soft-band image on 2005 August
31. The position of the transient source is very close to the edge of the detector and
its image is truncated. In order to constrain the position as precisely as possible, we
took the following procedure. Throughout this paper, we used ray-tracing software
designed for the Suzaku telescopes (\texttt{xissim}; \cite{ishisaki07}) to simulate XIS
images. Simulation were conducted only for XIS1; the source is quite soft and most of
the events were recorded in XIS1.

First, we registered the astrometry of the XIS image by matching the position of
1E\,0102.2--7219 with that of Chandra. The astrometry of the Chandra image is accurate
to $\sim$0\farcs6\footnote{See http://asc.harvard.edu/proposer/POG/ for details.}. We
simulated an XIS image based on the Chandra image (obsID\,$=$\,5139), and constructed
R.\,A. and Dec. projections of the surface brightness in the 0.2--2.0~keV band. The
observed profile was fitted to the simulated profile by shifting 13\arcsec\ and
19\arcsec\ in the R.\,A. and Dec. directions, respectively. The uncertainty of the
fitting is estimated to be $\sim$1\arcsec\ in both directions, which is inherited to the
uncertainty of the position determination of the transient source.

\begin{figure}
 \begin{center}
  \FigureFile(80mm,80mm){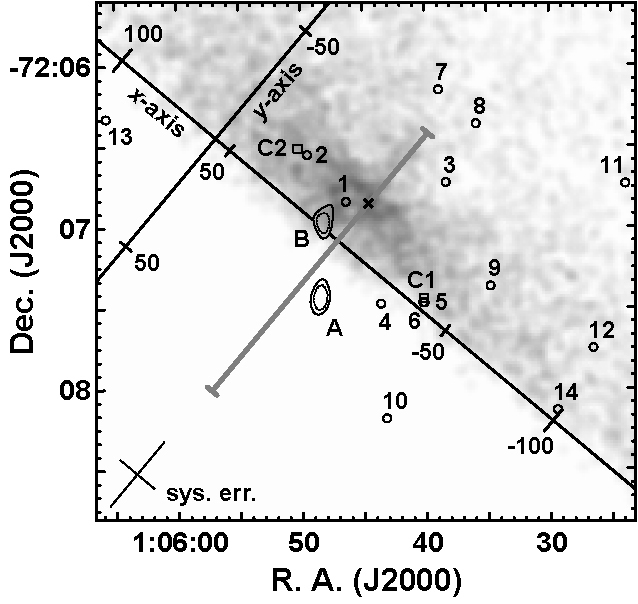}
 \end{center}
 \caption{
 Close-up image of the transient source in gray scale. The observed peak position is
 shown by a cross. The open circles are the positions of the nearby sources with labels
 given in table~\ref{tb:t3}, while the open squares are the positions of the two faint
 sources (C1 and C2; also in table~\ref{tb:t3}) found in the archived Chandra
 images. The range of initial positions of the simulated source is given by the gray
 line. The contours indicate 90\% and 99\% confidence ranges of the transient source
 position. The systematic uncertainty of the position determination is indicated as a
 cross at the bottom left.}\label{fg:f2}
\end{figure}

Second, we simulated XIS images of a point source at various positions toward the inside
and outside of the FoV from the observed peak of the source. Figure~\ref{fg:f2} shows
the layout of the simulated positions. We define the axes parallel and perpendicular to
the detector edge as $x$ and $y$, respectively. The origin of the $x$-axis is defined so
that the observed peak of the transient source is 0. The origin of the $y$-axis is
placed at the field edge, and negative and positive values of $y$ indicate that the
position is inside and outside of the field, respectively. The unit of the coordinate is
pixels. The apparent peak of the source is at ($x$, $y$)\,$=$\,(0, $-$17).

We simulated XIS images at positions along the line between ($x$, $y$)\,$=$\,(0, $-$50)
and (0, 70) with a step of $\Delta y$\,$=$\,1 (the gray line in figure~\ref{fg:f2}). For
the simulated two-dimensional image at each position, we constructed a one-dimensional
surface brightness profile inside the FoV of the detector to match with the observed
profile. The profiles were projected on the $x$-axis and the counts in the range of
$y$~$>$~--183 were accumulated, so that it covers the entire extent of the observed
image. In the fitting, the simulated profile was allowed to move along the $x$-axis, so
that its peak matches with that of the observed profile. 

With the $x$-projected profile, we can constrain both the $x$ and $y$ positions of the
transient source. On one hand, the peak position of the profile is sensitive to the
assumed $x$ position of the transient source. On the other hand, the width of the
profile is sensitive to the assumed $y$ position; this is because the width of the
simulated profile becomes narrower as the assumed $y$ position moves away from the
edge. We do not use the $y$-projected profiles as their peaks and widths are insensitive
to the assumed position of the source.

\begin{figure*}
 \begin{center}
  \FigureFile(160mm,160mm){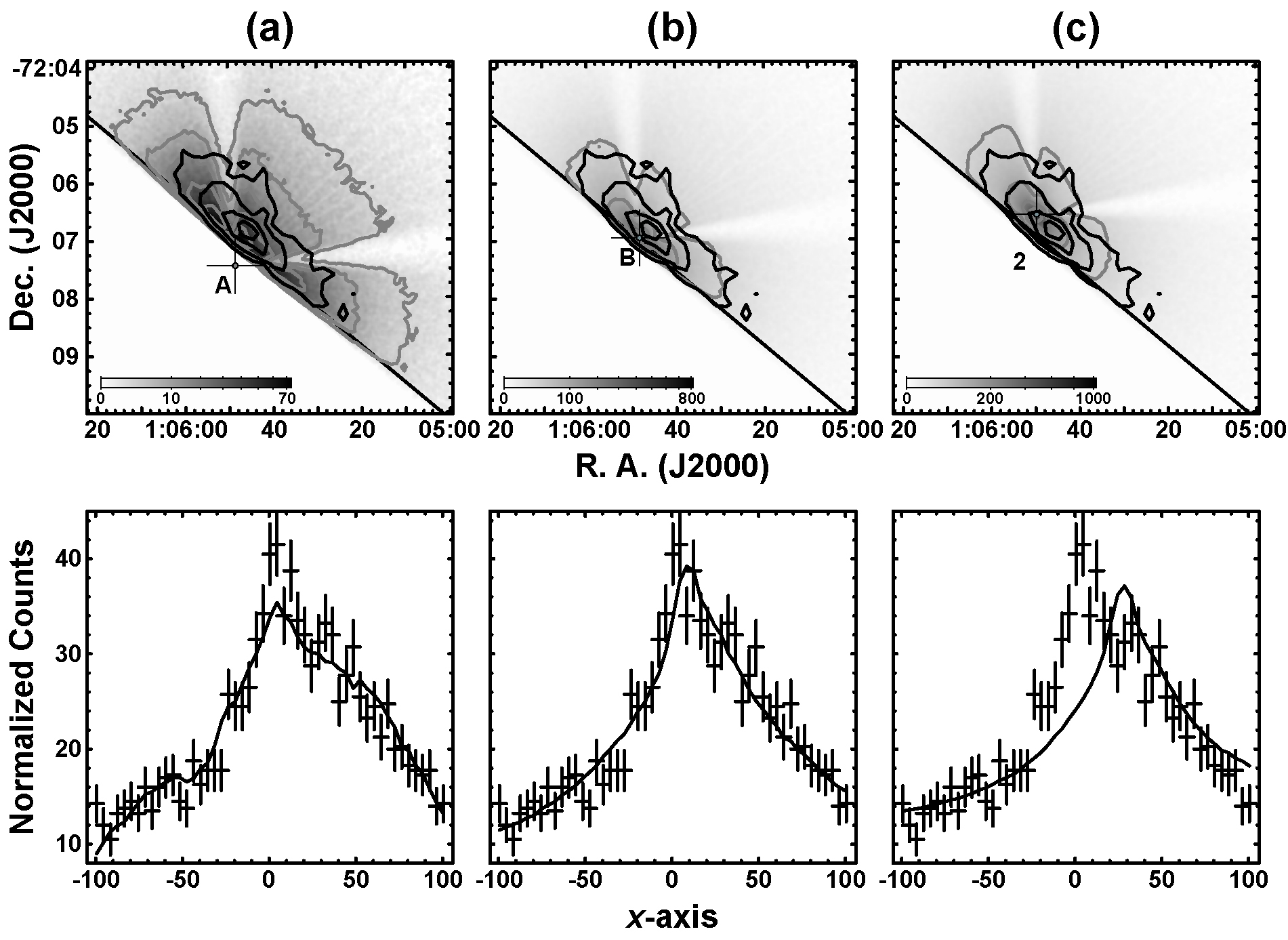}
 \end{center}
 \caption{
 Simulated images and profiles at three different positions; (a) the local minimum A, (b) local minimum B, and (c) the second closest source in
 table~\ref{tb:t3}. The upper panels show the simulated images in gray scale. The
 intensity of the simulated and observed images in an arbitrary unit is shown with gray
 and black contours, respectively. The assumed positions are shown with pluses. The
 lower panels show the projected profiles of the simulated (lines) and the observed
 (pluses) images.}\label{fg:f3}
\end{figure*}

After this procedure for every point along the gray line in figure~\ref{fg:f2}, we
obtained the 90\% and 99\% confidence ranges of the source position, which are shown as
contours in figure~\ref{fg:f2}. We have two islands of local minimum; one (A in
figure~\ref{fg:f2}) is outside of the FoV at ($x$, $y$)\,$=$\,($-$8.5, 20) or (R.\,A.,
Dec.)\,$=$\,(\timeform{01h05m49s}, \timeform{-72D07'26''}), and the other (B in
figure~\ref{fg:f2}) is inside of the FoV at ($x$, $y$)\,$=$\,(8.5, $-$2) or (R.\,A.,
Dec.)\,$=$\,(\timeform{01h05m48s}, \timeform{-72D06'57''}). The former has the smaller
$\chi^{2}$ value of the fit. The simulated images and the best-fit profiles at the two
local minima are shown in figure~\ref{fg:f3} (a) and (b).

The systematic uncertainty of this procedure was assessed using a different XIS
observation of 0836$+$714. A bright source is found in the image close to the edge but
well within the detector. We artificially truncated the image of the source and derived
the displacement of the real and reconstructed peak positions. The derived uncertainty
of $\sim$8\arcsec\ and $\sim$15\arcsec\ respectively for the $x$ and $y$ directions is
shown as a cross at the bottom left in figure~\ref{fg:f2}. We designate the source
Suzaku\,J0105--72.

\subsection{Temporal Analysis}
\begin{figure}
 \begin{center}
  \FigureFile(80mm,80mm){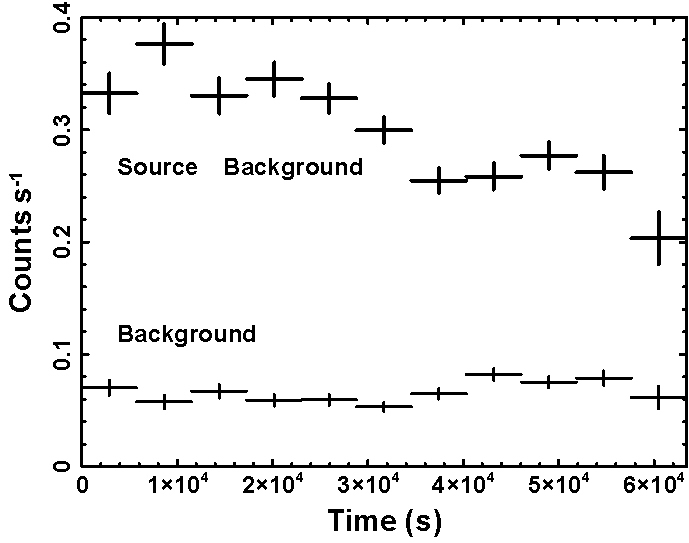}
 \end{center}
 \caption{
 Background-subtracted source (thick pluses) and background (thin pluses) light curves
 with a binning size of 5760~s, the period of the Suzaku orbit. Events taken with the
 four XIS in 0.2--2.0~keV were merged. The origin of the time is the start time of the
 observation.}\label{fg:f4}
\end{figure}

We constructed a background-subtracted light curve of the transient source
(figure~\ref{fg:f4}). The source and background photons were accumulated from the solid
and dashed circles truncated by the field edge in figure~\ref{fg:f1}. The source circle
around the observed peak has a radius of 3\arcmin, which would include $\sim$90\% of the
photons from a point source if the whole region is inside the FoV. The background region
was selected from a region devoid of X-ray emission at the edge to have the same radius
and the same truncated fraction with the source region.

We found a decline of the count rate in the light curve. We fitted the curve with a
constant count rate model, which was rejected by a $\chi^{2}$ test with a
$\chi^{2}/\rm{d.o.f}$ value of $\sim$93$/$10. No such trend was found in the background
light curve, which indicates that the decline is intrinsic to the transient source.

We also searched for any periodicity of signals using the Lomb method \citep{lomb76} and
the epoch folding search technique. We found no significant periodicity, except for
those related to the Suzaku orbit.

\subsection{Spectral Analysis}
\begin{figure}
 \begin{center}
  \FigureFile(80mm,80mm){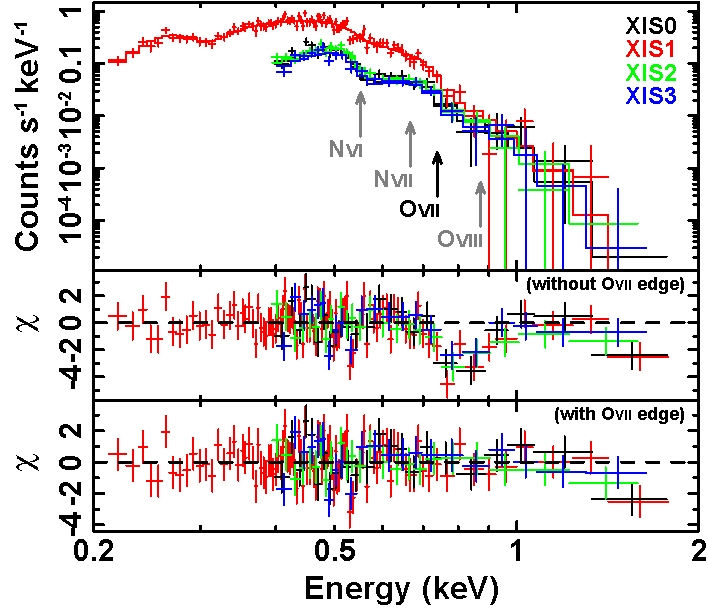}
 \end{center}
 \caption{
 XIS spectrum and the best-fit model. The background-subtracted spectra are shown with
 pluses in the upper panel in different colors for the four censors (black, red, green,
 and blue for XIS 0--3). The best-fit model (absorbed blackbody with an edge) is shown
 with solid lines. The energies of the K edges are indicated by arrows. The lower two
 panels show the residuals from the absorbed blackbody models with and without the
 O\emissiontype{VII} edge.}\label{fg:f5}
\end{figure}

We constructed the background-subtracted spectrum in 0.4--2.0~keV for the FI chips and
0.2--2.0~keV for the BI chip (figure~\ref{fg:f5}). The source and the background signals
were integrated from the same regions with the temporal analysis. The spectrum is very
soft with almost all photons below 2~keV and shows no conspicuous emission lines. In
order to fit the spectrum, we created the detector and mirror responses on the
observation date at two local minima (A and B in figure~\ref{fg:f2}) using the
\texttt{xisrmfgen} and \texttt{xissimarfgen} tools. The spectra and responses were
constructed independently for the four CCDs and the fitting trials were conducted
simultaneously for the set of four spectra. In order to account for the possible
normalization differences among the four CCDs, we added normalization parameters
relative to XIS1.

We fitted the spectrum with several continuum (blackbody, bremsstrahlung, and power-law)
models with interstellar extinction. The blackbody model with a temperature of
\kt~$\sim$72~eV and the interstellar extinction of
\nh~$\sim$4.9$\times$10$^{20}$~H~atoms~cm$^{-2}$ yielded the best-fit result, which
explains the global spectral shape quite well. The amount of interstellar extinction is
consistent with the value toward the SMC \citep{dickey90}. However, we see an edge-like
residual at $\sim$0.74~keV as shown in the middle panel of figure~\ref{fg:f5}. We
therefore modified the blackbody model by adding an edge model with two more free
parameters, the edge energy ($E$) and the optical depth at the edge ($\tau$), and
obtained an improved acceptable fit. Based on the F-test, we found that the improvement
is statistically significant ($<$10$^{-6}$ of a chance probability).

The best-fit parameters are given in table~\ref{tb:t2} separately for the position A and
B. X-ray flux (\fx) and luminosity (\lx) are derived in the 0.2--2.0~keV band. A
distance of 60~kpc is assumed for the luminosity. The best-fit model is given in
figure~\ref{fg:f5} for the position A. The parameters for the spectral shape (\nh, \kt,
$E$, and $\tau$) do not differ so much between the position A and B. This is because the
half power diameter and the vignetting function of the telescopes do not depend on
energy at $<$5~keV. On the other hand, the parameters for the normalization (\fx) are
different. This is because A is further away from the observed peak, thus requires a
brighter intrinsic flux.

\begin{table*}
 \begin{center}
 \caption{Best-fit parameters of the fitting model}\label{tb:t2}
  \begin{tabular}{lllll}
   \hline\hline
   Components & Par.                        & Units                    & Values (at pos.~A)\footnotemark[$*$]       & Values (at pos.~B)\footnotemark[$*$] \\
   \hline
   Absorption & \nh                         & (cm$^{-2}$)  & 4.87 $_{-0.42}^{+0.44}$~$\times$10$^{20}$  & 4.86 $_{-0.42}^{+0.44}$ $\times$10$^{20}$ \\
   Blackbody  & \kt                         & (eV)                     & 71.6 $_{-2.0}^{+2.1}$                      & 71.6 $_{-2.0}^{+2.1}$ \\
              & \fx\footnotemark[$\dagger$] & (erg~s$^{-1}$~cm$^{-2}$) & 1.02 $_{-0.06}^{+0.04}$~$\times$10$^{-11}$ & 4.66 $_{-0.45}^{+0.31}$ $\times$10$^{-12}$ \\
              & \lx\footnotemark[$\dagger$] & (erg~s$^{-1}$)           & 1.35 $_{-0.07}^{+0.05}$~$\times$10$^{37}$  & 6.14 $_{-0.30}^{+0.37}$ $\times$10$^{36}$ \\
   Edge       & $E$                         & (keV)                    & 0.74 $_{-0.02}^{+0.02}$                    & 0.74 $_{-0.02}^{+0.02}$ \\
              & $\tau$                      &                          & 1.22 $_{-0.39}^{+0.52}$                    & 1.22 $_{-0.39}^{+0.43}$ \\
   Constant   & XIS0                        &                          & 0.98 $_{-0.07}^{+0.07}$                    & 1.00 $_{-0.07}^{+0.07}$ \\
              & XIS1                        &                          & 1.00 (fixed)                               & 1.00 (fixed) \\
              & XIS2                        &                          & 0.81 $_{-0.05}^{+0.06}$                    & 0.94 $_{-0.06}^{+0.07}$ \\
              & XIS3                        &                          & 0.98 $_{-0.07}^{+0.08}$                    & 1.05 $_{-0.08}^{+0.08}$ \\
   \hline
   \multicolumn{3}{l}{$\chi^{2}/\rm{d.o.f}$ ($\chi^{2}_{\rm{red}}$)}   & 187.13/161 (1.16)                          & 187.29/161 (1.16) \\
   \hline
   \multicolumn{5}{@{}l@{}}{\hbox to 0pt{\parbox{130mm}{\footnotesize
        \footnotemark[$*$] The uncertainties are for the 90\% confidence ranges, which
   do not include systematic uncertainties in the instrumental calibration. See
   http://www.astro.isas.jaxa.jp/suzaku/doc/suzaku\_td/ suzaku\_td.html for details.
	\par\noindent
        \footnotemark[$\dagger$] Flux and luminosity are estimated in a range of 0.2--2.0~keV.
        The distance to the SMC (60~kpc) is assumed to derive the luminosity.
   }\hss}}
  \end{tabular}
 \end{center}
\end{table*}

\section{Discussion}
\subsection{Counterpart Search and Long-term Behavior}
First, in search for a possible bursts of this source in the past, we retrieved all the
available archived data taken in the vicinity by Einstein, ROSAT, ASCA, Beppo-SAX,
Chandra, and XMM-Newton. We inspected a total of 133 observation that span $\sim$28
years with an integrated exposure time of $\sim$2~Ms. No source was found in the error
region at a comparable brightness with the Suzaku source.

Next, in search for the quiescent X-ray emission of this source and the counterpart in
other wavelengths, we further retrieved the SIMBAD database as well as the published
X-ray source lists using Einstein (\cite{inoue83}; \cite{seward81}; \cite{wang92}),
ROSAT (\cite{kahabka99a}; \cite{haberl00}; \cite{sasaki00}), and ASCA
\citep{yokogawa03}. Table~\ref{tb:t3} lists the sources within 2\arcmin\ of the observed
peak of the Suzaku source.

We supplemented the list with two faint X-ray sources, which we found during the visual
inspection of the archived images. These two sources appear most significantly in a
series of Chandra images. We determined their positions, named them
CXOU\,J010540.1--720726 and J010550.3--720631, and refer to them as C1 and C2
hereafter. Four X-ray sources (sources 5, 6, C1, and C2) are listed in
table~\ref{tb:t3}. From the positional coincidence, we consider that the two ROSAT
sources (sources 5 and 6) and a Chandra source (C1) are the same.

Among the sources in table~\ref{tb:t3}, source 1 is the only likely counterpart of the
Suzaku source for its positional coincidence including the systematic uncertainty. Other
sources including the two X-ray sources are displaced too far from both the islands A
and B, thus are unlikely to be the counterpart. At the position of source 2, the closest
source except for source 1, the simulated X-ray profile is inconsistent with the
observed one (figure~\ref{fg:f3}c).

\begin{figure}
 \begin{center}
  \FigureFile(80mm,80mm){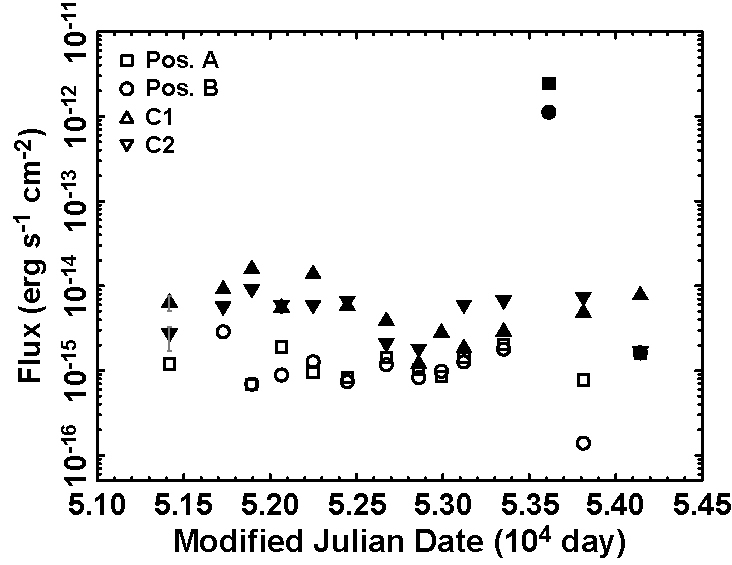}
 \end{center}
 \caption{
 Long-term flux (0.5--2.0~keV) variation of the Suzaku and the neighboring Chandra
 sources. The Suzaku transient flux at the positions A and B are plotted by filled
 squares and circles, respectively. The 3~$\sigma$ upper limits at these positions by
 Chandra are also plotted with open symbols. The flux of the two Chandra sources (C1 and
 C2) are shown with upward and downward triangles, respectively. The typical uncertainty
 for these sources are given for the first data points. Chandra data within ten days
 were grouped.}\label{fg:f6}
\end{figure}

\begin{figure}
 \begin{center}
  \FigureFile(80mm,80mm){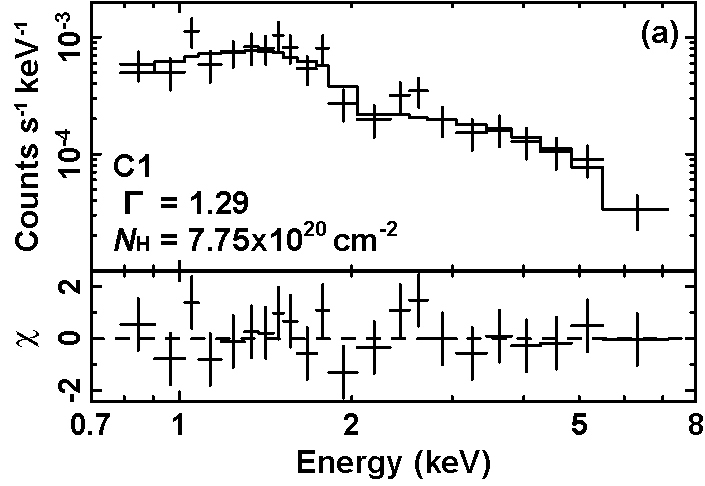}
  \FigureFile(80mm,80mm){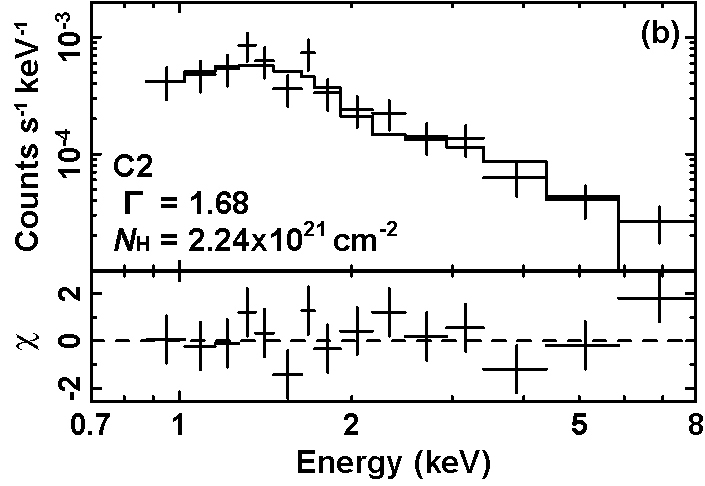}
 \end{center}
 \caption{
 ACIS spectra of the two Chandra sources (C1 and C2). The upper panels show the
 background-subtracted spectra combining all the data sets (pluses) and the best-fit
 absorbed power-law models (solid lines). The lower panels show the residuals to the
 fit.}\label{fg:f7}
\end{figure}

To reinforce our claim that the Suzaku source is different from the two nearby X-ray
sources, we reduced 26 data sets taken by the Advanced CCD Imaging Spectrometer (ACIS;
\cite{garmire03}) onboard Chandra \citep{weisskopf02} to reveal their nature. The
exposure times of all ACIS observations were too short to construct spectra. We derived
the photometric flux as the mean energy times the net count rate divided by the
accumulation area \citep{tsujimoto05} in the soft-band at each epoch and found that both
C1 and C2 are faint persistent sources (figure~\ref{fg:f6}). By combining all the data
sets, we constructed composite X-ray spectra (figure~\ref{fg:f7}). The two sources show
very similar spectra with hard and featureless emission. The absorbed power-law model
well explains these spectra. The best-fit index of power, 0.5--2.0~keV flux, and the
amount of extinction are 1.3--1.7, 4--5$\times$10$^{-15}$~erg~s$^{-1}$~cm$^{-2}$, and
0.8--2.2$\times$10$^{21}$~cm$^{-2}$, respectively. The hard spectra extending beyond
$\sim$5~keV and the amount of extinction larger than that of the Suzaku source comprise
a sharp contrast with the Suzaku spectrum, indicating that C1 and C2 are not the Suzaku
counterpart. We examined a Digitalized Sky Survey (DSS) image at C1 and C2 and found no
optical source. From all these properties, both C1 and C2 are likely to be background
active galactic nuclei.

Using the same Chandra data sets, we can derive the flux upper limits at the Suzaku
position for the last seven years and a half. Figure~\ref{fg:f6} shows a long-term trend
of the Suzaku and the two Chandra sources. The figure indicates that the transient
source experienced a burst during the Suzaku observation, and the flux was amplified by
more than 10$^{3}$ times compared to the undetected quiescent level.

\begin{table*}
 \begin{center}
 \caption{List of nearby sources}\label{tb:t3}
  \begin{tabular}{rlcccll}
   \hline\hline
   ID & Source name\footnotemark[$*$] & R.\,A.    & Dec.      & Separation\footnotemark[$\dagger$] & Object types\footnotemark[$*$] & References\footnotemark[$\ddagger$] \\
      &                               & (J2000.0) & (J2000.0) &                                    &                                & \\
   \hline
    1 & 2dFS\,2064               & \timeform{01h05m46s} & \timeform{-72D06'51''} & \timeform{0.14'} & Star (B0) & [1] \\
    2 & MA93\,1554               & \timeform{01h05m50s} & \timeform{-72D06'33''} & \timeform{0.48'} & Emission-line star & [2] \\
    3 & FBR2002\,J010538--720643 & \timeform{01h05m38s} & \timeform{-72D06'43''} & \timeform{0.50'} & Radio source & [3] \\
    4 & 2dFS\,2059               & \timeform{01h05m44s} & \timeform{-72D07'28''} & \timeform{0.63'} & Star (F0) & [1] \\
    5 & RX\,J0105.7--7207        & \timeform{01h05m40s} & \timeform{-72D07'28''} & \timeform{0.70'} & X-ray source & [4] \\
    6 & HFP2000\,135             & \timeform{01h05m40s} & \timeform{-72D07'28''} & \timeform{0.70'} & X-ray source & [5] \\
    7 & MA93\,1542               & \timeform{01h05m39s} & \timeform{-72D06'09''} & \timeform{0.83'} & Emission-line star & [2] \\
    8 & OGLE\,SMC--SC10\,106764  & \timeform{01h05m36s} & \timeform{-72D06'22''} & \timeform{0.83'} & Star & [6] \\
    9 & OGLE\,SMC--SC10\,104288  & \timeform{01h05m35s} & \timeform{-72D07'22''} & \timeform{0.91'} & Unknown & [7] \\
   10 & MA93\,1548               & \timeform{01h05m43s} & \timeform{-72D08'11''} & \timeform{1.34'} & Emission-line star & [2] \\
   11 & 2dFS\,2011               & \timeform{01h05m24s} & \timeform{-72D06'44''} & \timeform{1.59'} & Star (B5) & [1] \\
   12 & MA93\,1531               & \timeform{01h05m27s} & \timeform{-72D07'45''} & \timeform{1.65'} & Emission-line star & [2] \\
   13 & 2dFS\,2110               & \timeform{01h06m06s} & \timeform{-72D06'20''} & \timeform{1.70'} & Emission-line star & [1] \\
   14 & IRAS\,F01038--7224       & \timeform{01h05m29s} & \timeform{-72D08'08''} & \timeform{1.73'} & Infra-red source & [8] \\
   \hline
   C1 & CXOU\,J010540.1--720726  & \timeform{01h05m40s} & \timeform{-72D07'26''} & \timeform{0.68'} & X-ray source & this work \\
   C2 & CXOU\,J010550.3--720631  & \timeform{01h05m50s} & \timeform{-72D06'31''} & \timeform{0.55'} & X-ray source & this work \\
   \hline
   \multicolumn{7}{@{}l@{}}{\hbox to 0pt{\parbox{170mm}{\footnotesize
        \footnotemark[$*$] The names and object types follow the SIMBAD designation except for C1 and C2.
        The spectral types are given for stars in the object type when they are available.
        \par\noindent
	\footnotemark[$\dagger$] Separation from the observed peak.
        \par\noindent
        \footnotemark[$\ddagger$] [1] \citet{evans04}, [2] \citet{meyssonnier93}, [3] \citet{filipovi02}, [4] \citet{filipovi00},
        [5] \citet{haberl00}, [6] \citet{udalski98}, [7] \citet{ngeow06}, [8] \citet{moshir90}.
        }\hss}}
  \end{tabular}
 \end{center}
\end{table*}

\subsection{Nature of Suzaku\,J0105--72}
The transient source has a very soft spectrum of a blackbody temperature of $\sim$72~eV
(figure~\ref{fg:f5}). Assuming a distance of 60~kpc, the bolometric luminosity is
$\sim$10$^{37}$~erg~s$^{-1}$ and the blackbody sphere has a radius of
$\sim$10$^{8}$~cm. The light curve (figure~\ref{fg:f4}) shows the gradual decline of the
X-ray flux. The long-term behavior (figure~\ref{fg:f6}) indicates that the source was in
a burst with the amplified flux of more than 10$^{3}$ times than the undetected
quiescent level.

All these properties are commonly seen among supersoft sources (SSSs). SSSs are
considered to be a binary of a white dwarf and a companion star \citep{heuvel92} in the
SSS phase, in which mass accretion from the companion star fuels hydrogen burning on the
white dwarf surface. The X-ray spectra observed by CCD or proportional counter
spectrometers are fitted by a blackbody model of \kt~$\lesssim$~100~eV and the
bolometric luminosity of 10$^{36}$--10$^{38}$~erg~s$^{-1}$ \citep{greiner00}. We
therefore conclude that the nature of the transient source is a white dwarf binary,
which was in the declining end of the SSS phase during the Suzaku observation.

SSSs show a variety of temporal behaviors. Some sources are constant supersoft X-ray
emitters fueled by continuous mass inflow and steady nuclear burning (e.g., CAL\,87;
\cite{heuvel92}) with occasional off-states (e.g, CAL\,83; \cite{kahabka98}). Others
have eruptive nature recognized as classical novae (e.g., RX\,J0513.9--6951;
\cite{reinsch99}), which can be recurrent (e.g., RS\,Oph; \cite{hachisu07}). The
transient nature of the Suzaku\,J0105--72 indicates that the source is in the latter
group.

The possible optical counterpart, source 1, is a spectroscopically-identified B0
star. This source may be the companion star of the Suzaku source. We caution, however,
that many anonymous DSS sources can be found in the positional confidence range of the
Suzaku source. Follow-up studies are necessary to identify the companion star.

\subsection{White Dwarf Mass}
Although the observational data are limited, we can obtain a rough estimate of the mass
of the white dwarf. Two independent estimates can be derived from the X-ray light curve
and the plasma temperature, both of which indicate that the mass is $\approx$1.2~$M_{\odot}$.

First, the duration of the supersoft phase is a function of mass (figure~3 in
\cite{hachisu06}). The supersoft phase is seen at a later stage of bursts in classical
novae. In the optically thick wind model (\cite{hachisu06} and references therein), the
supersoft phase emerges when the wind stops and the inflated photosphere shrinks after
the optical burst and ends when the hydrogen shell burning is extinguished. As the mass
increases, the duration of the supersoft phase becomes shorter. This is because for a
larger mass, hence for a stronger surface gravity, the accreting mass can be smaller to
ignite the thermonuclear process, resulting in a shorter time scale to exhaust the fuel
\citep{hachisu05}.

In the light curve of the supersoft burst (figure~\ref{fg:f6}), the Chandra observation
$\sim$200 days after the Suzaku burst gives a strong flux upper limit at the source
position. Besides, we do not find any significant emission at the position in the Suzaku
image obtained $\sim$110 days after the burst on 2005 December 16 (table~\ref{tb:t1}),
despite the facts that the image was exposed for three times longer and that the
observation has a roll angle to cover the source well within the XIS FoV. These suggest
that the supersoft phase faded out quickly in less than $\sim$3.5 month. We compared the
duration with the calculation for the smallest metallicity value (figure~3 and table~8
in \cite{hachisu06}) and found that the mass is estimated to be larger than
$\sim$1.2~$M_{\odot}$. The rapid decay is reminiscent to the recent supersoft burst from
the recurrent nova RS\,Oph \citep{osborne06}, which faded out in $\sim$60 days.

Second, the effective temperature of the soft X-ray blackbody emission is also a
function of mass (figure~13 in \cite{ebisawa01}), in which the temperature becomes
larger as the mass increases. This is because for a larger mass, hence for a stronger
surface gravity, the plasma electron density increases, which suppresses the
ionization. To achieve the observed level of ionization, the plasma temperature has to
be higher \citep{ebisawa01}. We observed a color temperature of 72$\pm$2~eV, which we
assume is similar to the effective temperature. The mass is estimated to be
$\sim$1.0--1.2~$M_{\odot}$ for the plasma at a local thermal equilibrium.

\subsection{Absorption Edge}
Some SSSs show absorption features in their blackbody spectra, which is thought to arise
from the absorption by the atmosphere of the white dwarf (\cite{heise94};
\cite{teesellng96}; \cite{hartmann97}). These features provide valuable tools to
diagnose the physical status of the atmospheric plasma.

The Suzaku spectrum shows a conspicuous edge feature at $\sim$0.74~keV with an optical
depth of $\sim$1.2 (figure~\ref{fg:f5} and table~\ref{tb:t2}). We consider that this is
the O\emissiontype{VII} K edge at 0.7393~keV \citep{scofield01}. In contrast, we do not
see significant features of other K absorption edges by O\emissiontype{VIII} at
0.8714~keV, N\emissiontype{VI} at 0.5521~keV, and N\emissiontype{VII} at 0.6671~keV
\citep{scofield01}. The 90\% confidence upper limits to the optical depth of these
features were measured to be $\sim$1.18, $\sim$0.06, and $\sim$0.25, respectively.

The presence of the O\emissiontype{VII} K edge and the upper limit of the
O\emissiontype{VIII} K edge set an upper limit to the temperature of the
atmosphere. \citet{ebisawa01} calculated the fractional density of ions at various
ionization stages for several elements (C, N, O, Ne, and Fe), as a function of plasma
temperature (0--150~eV), and two electron densities ($n_{e}$\,$=$\,10$^{18}$ and
10$^{19}$~cm$^{-3}$). The K edge absorption cross sections for O\emissiontype{VII} and
O\emissiontype{VIII} are $\sim$2.4 $\times$10$^{-19}$ and $\sim$1.2
$\times$10$^{-19}$~cm$^{2}$, respectively \citep{lang06}. The ratio of the optical depth
($<$1.42) is converted to the ratio of the column density ($<$2.84) of
O\emissiontype{VIII} to O\emissiontype{VII}. From the ratio and figure~1 in
\citet{ebisawa01}, the atmospheric plasma temperature is constrained to be $\lesssim$58
and $\lesssim$68~eV for the electron density of 10$^{18}$ and 10$^{19}$~cm$^{-3}$,
respectively.

The lack of N absorption edges does not necessarily indicate that the nitrogen is
deficient in terms of the N/O ratio from the solar abundance. The metallicity of the SMC
is deficient from the solar value by several times \citep{russell92}, which we canceled
by using the ratio N/O. From the upper limits of the N\emissiontype{VI} K edge, the
relative abundance of N/O is $\lesssim$0.46 assuming \kt\,$=$\,60~eV and
$n_{e}$\,$=$\,10$^{18}$~cm$^{-3}$, which does not contradict with the solar value of
$\sim$0.14 \citep{anders89}.

Finally, we mention that the blackbody plus edge model may be an
oversimplification. Using a grating X-ray spectrometer, \citet{orio04} showed that the
spectrum of a SSS (CAL87) has in fact a complex of emission lines and absorption edges,
which appears as a blackbody spectrum in a study using a CCD spectrometer with a lower
resolution by $\sim$10~times \citep{ebisawa01}. Higher resolution spectroscopy
observations need be conducted in the future to confirm these results.

\section{Summary}
Using the Suzaku XIS, we detected a transient source during a routine calibration
observation of 1E\,0102.2--7219. The source was detected at the edge of the XIS
image. We constrained its position to a $\sim$20\arcsec\ accuracy with the aid of
ray-tracing simulations and named it Suzaku\,J0105--72.

No such transient source was found in the catalogued and archived data sets despite a
large number of observations, indicating that the burst event is quite rare. The source
showed a two-fold decline in flux during the Suzaku observation, probably at the
declining phase of a rare burst in which the flux was amplified by at least 10$^{3}$
times of the undetected quiescent level.

The spectrum is explained by a blackbody emission with a temperature of $\sim$72~eV. The
measured extinction of $\sim$4.9$\times$10$^{20}$~cm$^{-2}$ is typical to the
SMC. Assuming a distance of 60~kpc, the source has a bolometric luminosity of
$\sim$10$^{37}$~erg~s$^{-1}$.

From the X-ray temporal and spectral features, we concluded that Suzaku\,J0105--72 is a
SSS in the SMC. A short duration and a relatively high plasma temperature infer the
white dwarf mass of $\approx$1.2~$M_{\odot}$. The edge feature at $\sim$0.74~keV seen in
the XIS spectrum is probably due to O\emissiontype{VII} ions in the white dwarf
atmosphere. The presence of O\emissiontype{VII} K edge as well as the absence of
O\emissiontype{VIII}, indicate that the atmosphere has a plasma temperature of
$\lesssim$58--68~eV.

\bigskip

The authors thank the reviewer, Izumi Hachisu, for his insightful discussion. The Suzaku
data were obtained during the Science Working Group time. M.\,T. and M.\,M. are
financially supported by the Japan Society for the Promotion of Science. The Einstein,
ROSAT, ASCA, Beppo-SAX, Chandra, and XMM-Newton data were obtained through the High
Energy Astrophysics Science Archive Research Center Online Service provided by the
NASA/Goddard Space Flight Center. This research has made use of data obtained from Data
ARchives and Transmission System (DARTS), provided by PLAIN center at ISAS/JAXA, and of
the SIMBAD database, operated at CDS, Strasbourg, France.

\end{document}